# The Large Observatory For X-ray Timing: LOFT

**Enrico Bozzo[1]**

*ISDC – Department of Astronomy, University of Geneva*
*Chemin d'Ecogia 16, 1290, Versoix (Switzerland)*
*E-mail:* `enrico.bozzo@unige.ch`

**On behalf of the LOFT Consortium[2]**

LOFT, the Large Observatory for X-ray Timing, is a new space mission concept devoted to observations of Galactic and extra-Galactic sources in the X-ray domain with the main goals of probing gravity theory in the very strong field environment of black holes and other compact objects, and investigating the state of matter at supra-nuclear densities in neutron stars. The instruments on-board LOFT, the Large area detector and the Wide Field Monitor combine for the first time an unprecedented large effective area (~10 m$^2$ at 8 keV) sensitive to X-ray photons mainly in the 2-30 keV energy range and a spectral resolution approaching that of CCD-based telescopes (down to 200 eV at 6 keV). LOFT is currently competing for a launch of opportunity in 2022 together with the other M3 mission candidates of the ESA Cosmic Vision Program.

---

[1] Speaker

[2] Members of the LOFT Consortium are listed on the project official website: http://www.isdc.unige.ch/loft

# 1 Introduction

LOFT [1] is a new medium-size space mission being studied by the European space Agency since February 2011 within the context of the Cosmic Vision Program (2015-2025). The LOFT payload comprises of an Large Area Detector (LAD, [2]), and a Wide Field Monitor (WFM, [3]). The LOFT satellite (Figure 1-1) is currently being conceived to be launched into a low equatorial orbit by a Soyuz launcher. An updated table summarizing the anticipated performances of the LAD and WFM is provided on the mission web-site[3].

Among the different topics of the Cosmic Vision program, LOFT is specifically designed to investigate "matter under extreme conditions". By taking advantage of the large collecting area and fine spectral resolution of its instruments (see Sections 1.1 and 1.2), LOFT will exploit the technique of X-ray timing, combined with spectroscopic measurements, to probe for the first time General Relativity very close (a few gravitational radii, $r_g$) to black-holes (BHs) and provide unprecedented constraints on the equation of state (EoS) of supernuclear densities in neutron stars (NSs). A summary of the LOFT science is provided in Section 2.

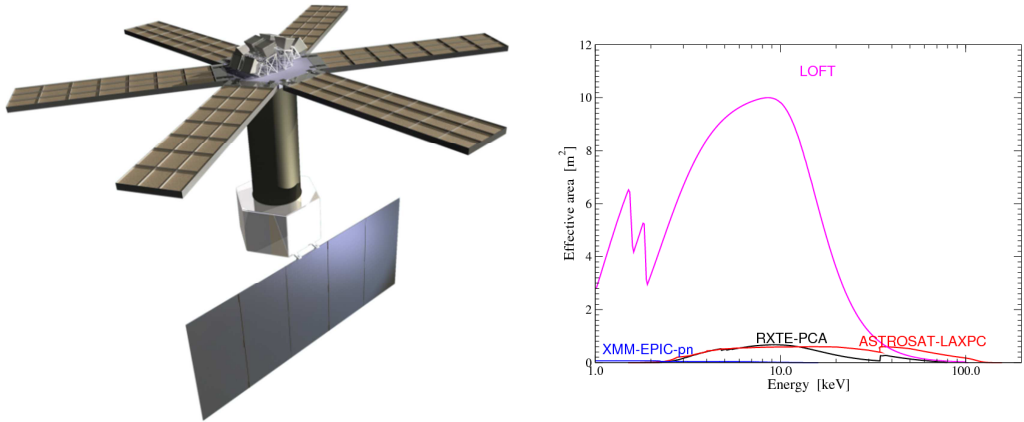

*Figure 1-1. Left: The LOFT Satellite. The LAD is composed by the 6 deployable panels covered by 2016 SDDs. The WFM, comprising 10 coded-mask cameras is located on the top. The satellite optical bench, bus and the solar panels are also displayed. Right: Plot of the LAD effective area as a function of energy. As a comparison a few other X-ray observatories are displayed on the same scale.*

## 1.1 The Large Area Detector

The LAD is a collimated non-imaging instrument performing pointed X-ray observations mainly in the 2-30 keV energy band. In the current LOFT design, it comprises of six panels organized in 21 modules each. A module hosts 16 Silicon Drift Detectors (SDDs), which provide a sensitive area for X-ray detection of about 76 cm$^2$. A total of 2016 detectors (i.e., 126 modules) are needed to complete the whole LAD, resulting in a total geometric area of about 18

---

[3] http://www.isdc.unige.ch/loft/index.php/instruments-on-board-loft.

m$^2$. This translates into an effective area for scientific observations reaching ~10 m$^2$ at 8 keV (see Figure 1-1). The main operating energy range of the LAD is 2-30 keV (the LAD "extended mode" energy range is 30-80 keV, see [1] for details). The nominal energy resolution is 260 eV. For about 40% of the events, an energy resolution of 200 eV can be achieved[4] ("single anode events", [1]). The LAD SDDs are equipped with lead-glass collimators, designed on the heritage of the micro-channel plates developed for the Bepi-Colombo mission [4]. These collimators limit the instrument field of view (FOV) to within ~1 degree and are able to efficiently stop X-ray photons coming from outside this FOV up to energies of ~30-40 keV. Higher energy photons from the Cosmic X-ray Background leaking through the collimators give the largest contribution to the LAD background. The unprecedented large effective area of the LAD produces a unique scientific throughput for any of the observed targets. A source with a flux of 1 Crab into the instrument energy band will generate about 240,000 cts/s.

### 1.1.1 The LAD background and sensitivity

The LAD background has been analyzed and computed using Monte Carlo simulations of a mass model of the whole LOFT spacecraft and all known radiation sources in the LOFT orbit. The results reported in Figure 1-2 below indicates that the anticipated LAD background over the full 2-30 keV energy band is compliant with the requirement of 10 mCrab, while being <5 mCrab in the most important energy band (2-10 keV)[3].

The background simulations show that the LAD background is dominated (>70%) by high energy photons of Cosmic X-ray Background (CXB) and Earth albedo "leaking" and scattering through the collimator structure, which becomes less efficient at high energies. These two radiation sources are very stable and then predictable. Though, the rotation of the spacecraft relative to the Earth during the orbital motion makes their viewing angle to change. Due to a different energy spectrum of the CXB and albedo components, the varying orientation produces a small orbital modulation of the background, which is however entirely due to a geometrical effect. This is the largest source of LAD background variaton. The effect of the other potentially varying source - particle induced background - is greatly reduced by the very stable particle environment offered by the equatorial orbit and by the fact that this component accounts for less than 6% of the overall background. Overall, the largest modulation of the total LAD background is estimated as ~10% (for comparison, the background of the RossiXTE/PCA could vary by more than 200% and it was largely due to the intrinsically variable particle-induced background) and can be effectively described by a geometrical model. The variation is slow (orbital timescale) and smooth as it depends only on the varying viewing angles of Earth and open sky. In Figure 1-2 (top left) we show the results of a study in which the behavior of each background component was modelled along the satellite orbit.

---

[4] These values are all given at the end-of-life, i.e., when the environmental radiation damages on the SDDs during the mission lifetime are taken into account [1].

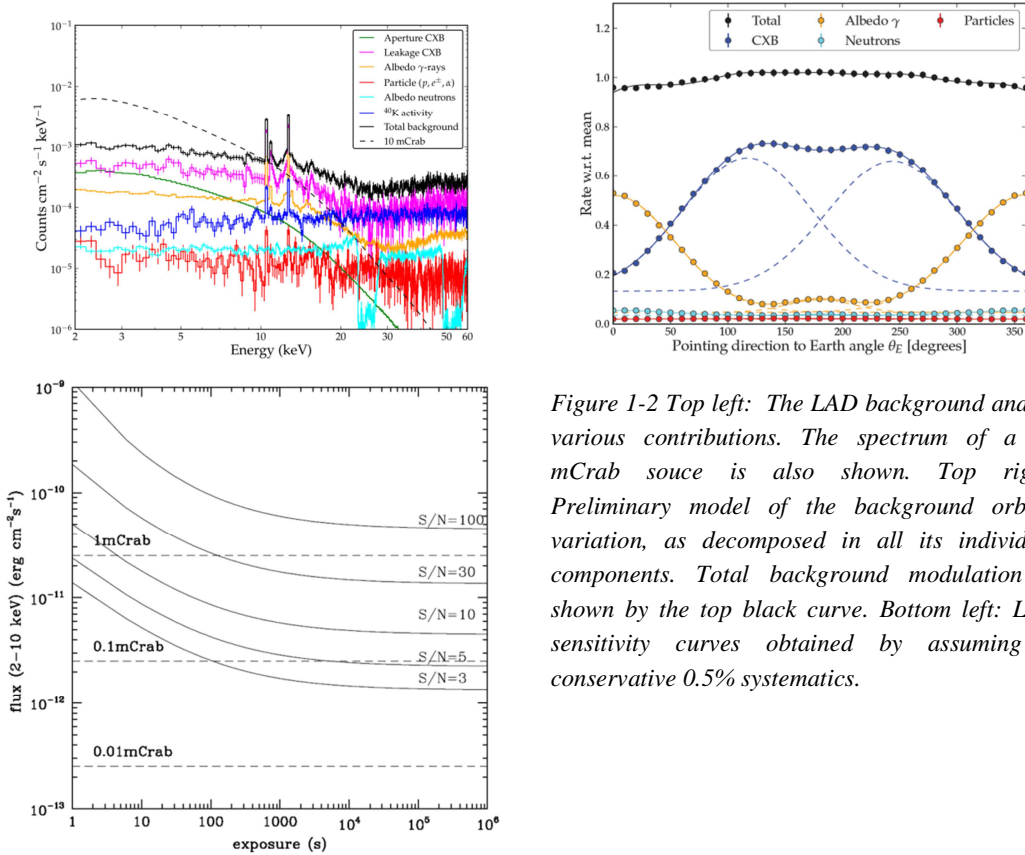

*Figure 1-2 Top left: The LAD background and its various contributions. The spectrum of a 10 mCrab souce is also shown. Top right: Preliminary model of the background orbital variation, as decomposed in all its individual components. Total background modulation is shown by the top black curve. Bottom left: LAD sensitivity curves obtained by assuming a conservative 0.5% systematics.*

A geometrical model, properly calibrated using in-orbit flat fields, is expected to predict the background at the level of 1% or better (2-10 keV), which is the LAD science requirement. Some of the LOFT science cases (especially the extragalactic science) will benefit from reaching a better background knowledge. For this reason, we introduced a "blocked collimator" (a collimator with the same stopping power but no holes) for one module of the LAD. This will continuously monitor all components of the LAD background, with the exception of the aperture background (about 10% or the total), accounting for 90% of the total background. In this way, in addition to the background modelling, the system will provide a continuous and independent benchmark of the background model. Preliminary simulations for different targets (i.e., different attitude configurations) show that an accuracy of ~0.1-0.3% in modelling the background variation can be achieved already over a few orbits timescale. This very low systematics permit to take advantage of the very high throughput (sensitivity) of the LAD up to >100 ks exposure (see Figure 1-2 bottom left and right).

Of course, in whatever science case the absolute value of the background becomes important, the local background could be predictable down to the cosmic variance level only (order of 0.7%). A better knowledge of the absolute level of the background will require local blank fields, to be requested in the observing proposal. An extensive discussion of the LAD background modelling is reported in [5].

## 1.2 The Wide Field Monitor

The WFM (Figure 1-3) is a coded-mask instrument whose main goal is to scan large fractions of the sky at once to catch interesting targets that can be followed up by the LAD. In the current design, the WFM comprises 10 cameras, each equipped with its own coded-mask and 4 SDDs similar to those employed for the LAD (reaching similar timing and spectral resolution but optimized for imaging purposes). The main operating energy range of the WFM is 2-50 keV. As the LOFT SDDs only provide arcmin position capabilities in one direction (the anode direction, see [1]), two orthogonal cameras are needed to achieve a unit with full 2D imaging capabilities [3]. The 5 units (10 cameras) of the WFM together achieve a FOV comprising roughly a rectangular region of 180°x90°, plus coverage in the anti-Sun direction (Figure 1-3). The WFM is endowed with on-board software able to identify and localize bright events (GRBs, magnetar flares, and other high-energy transients) with ~arcmin accuracy: the so-called LOFT Burst alert System. The position and trigger time of these events is broadcast to the ground using a VHF system to within 30 seconds from the discovery [3]. This will permit fast follow-up observations with those space and ground-based facilities that will be operating at the time of LOFT. For the brighter events the LBAS will collect and save data in full energy and timing resolution, to be sent to the ground through the main LOFT ground stations.

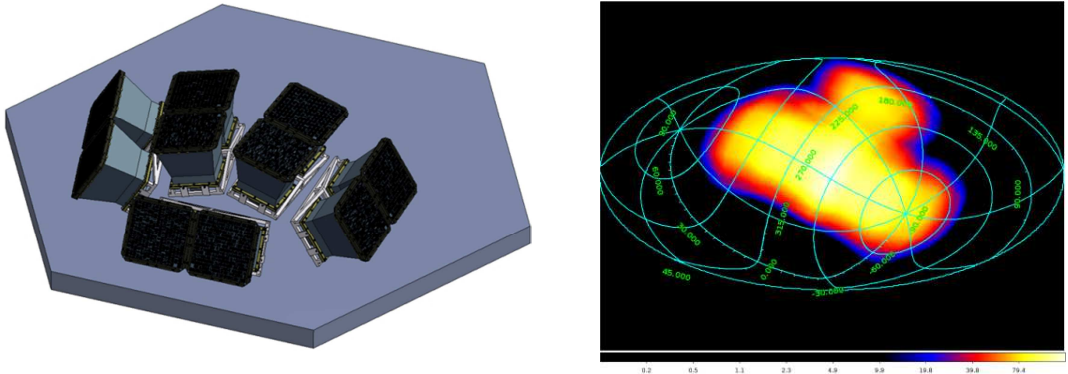

*Figure 1-3 Left: A sketch of the 10 cameras (5 units) of the WFM. Right: the WFM FOV in Galactic coordinates. The color code represents the effective area in each point in units of cm$^2$. In this example it is assumed that LOFT is pointing toward the Galactic center.*

## 2 LOFT Science

The capabilities of the instruments on-board LOFT will make it possible to constrain the NS EoS and test General Relativity in the strong field regime around BHs using different techniques. Two examples are given in Figure 2-1. A more exhaustive summary of the LOFT science is provided in [1]. The remaining observational time will be devoted to observatory science. Hundreds of X-ray sources can be observed with the LAD and the WFM, including Galactic binaries and AGNs. The exceptional throughput of the LAD will open for these sources new discovery windows, making fine spectroscopy and high resolution timing analyses accessible with unprecedented short integration times and high S/N. The wide FOV of the WFM will provide long term coverage for many X-ray sources and will permit to follow their behavior and catch any prompt impulsive event in a broad energy range and with a low energy threshold of ~2 keV.

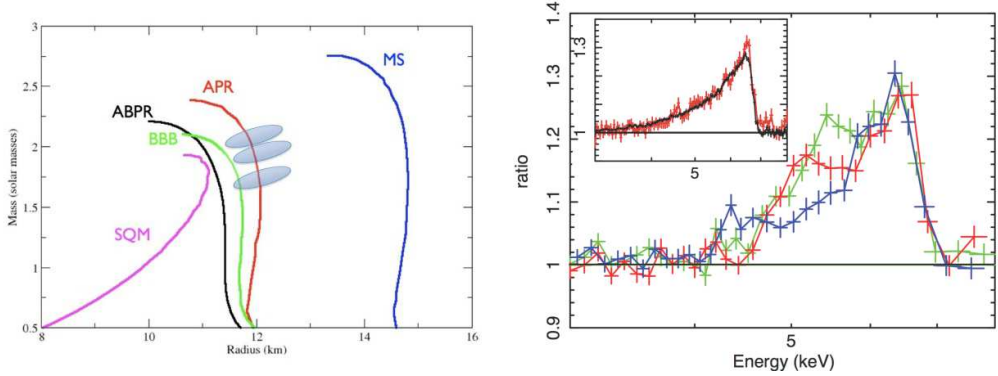

*Figure 2-1 Left: By exploiting different techniques LOFT will measure masses and radii of NSs with an accuracy of 5%. This will permit to obtain very high precision constraints on the dense matter EoS, and to distinguish even between EoS models that make similar predictions for the M-R relation. Right: Simulation of phase resolved spectroscopy of the Fe-K line from a hot spot orbiting on the accretion disk at $4r_g$, along the innermost stable circular orbit of a $2x10^7$ $M_\odot$ BH with spin a=0.5 in a 2.5 mCrab AGN. The hot spot lasts for 4 orbits (exposure time 24 ks). Line variations (between 4-7 keV) due to the orbiting spot (shown here as residuals after subtraction of the integrated spectrum) over 6 different phase intervals are clearly distinguishable (only the 3 brightest intervals are shown for clarity). The insert shows the 200 ks exposure integrated broad Fe-line profile for the XMM/EPIC-pn (red) and the LOFT/LAD (black). Fe K-line profiles probe the motion of matter close to NSs and BHs.*

**References**

[1] Feroci, M. 2012, Experimental Astronomy, 34, 415

[2] Zane, S. et al. 2012, SPIE, 8443, 2FZ

[3] Brandt, S. et al. 2012, SPIE, 8443, 2GB

[4] Fraser G. W. et al. 2010, Planetary and Space Science, 58, 79

[5] Campana, R. et al., 2013, Experimental Astronomy, in press (arXiv :1305.3789)